\documentclass[aps,prl,floatfix,twocolumn,superscriptaddress]{revtex4}

\pdfoutput=1
\usepackage{graphicx}
\usepackage{amssymb}
\usepackage{amsmath}

\newcommand {\ket}[1] {|#1 \rangle}
\newcommand {\bra}[1] {\langle#1 |}

\begin{document}

\title{Atomic three-body loss as a dynamical three-body interaction}

\date{28 October 2008}

\author{A. J. Daley}
\affiliation{Institute
for Theoretical Physics, University of Innsbruck, A-6020 Innsbruck, Austria\\ and Institute for Quantum Optics and Quantum Information of the
Austrian Academy of Sciences, A-6020 Innsbruck, Austria} 
\author{J. M. Taylor}
\affiliation{Department of Physics, Massachusetts Institute of Technology, 
Building 6C-411, Cambridge, MA 02139, USA}
\author{S. Diehl}
\affiliation{Institute
for Theoretical Physics, University of Innsbruck, A-6020 Innsbruck, Austria\\ and Institute for Quantum Optics and Quantum Information of the
Austrian Academy of Sciences, A-6020 Innsbruck, Austria} 
\author{M. Baranov}
\affiliation{Institute
for Theoretical Physics, University of Innsbruck, A-6020 Innsbruck, Austria\\ and Institute for Quantum Optics and Quantum Information of the
Austrian Academy of Sciences, A-6020 Innsbruck, Austria} 
\author{P. Zoller}
\affiliation{Institute
for Theoretical Physics, University of Innsbruck, A-6020 Innsbruck, Austria\\ and Institute for Quantum Optics and Quantum Information of the
Austrian Academy of Sciences, A-6020 Innsbruck, Austria} 

\begin{abstract}
We  discuss how large three-body loss of atoms in an optical lattice can give rise to effective hard-core three-body interactions. For bosons, in addition to the usual atomic superfluid, a dimer superfluid can then be observed for attractive two-body interactions. The non-equilibrium dynamics of preparation and stability of these phases are studied in 1D by combining time-dependent Density Matrix Renormalisation Group techniques with a quantum trajectories method. 
\end{abstract}

\maketitle

Cold atomic gases in optical lattices have proven a test bed for understanding novel quantum phases \cite{opticallattices} and non-equilibrium many-body dynamics \cite{vidal,tdmrg}. Recently, Syassen et al. \cite{moleculeloss1} showed that a strong two-body loss process for molecules in an optical lattice \cite{opticallattices} could produce an effective, elastic hard-core repulsion and thus a Tonks gas \cite{moleculeloss1,moleculeloss2}.  This is related to the quantum Zeno effect: a large loss dynamically suppresses processes creating two-body occupation on a particular site.  Whilst elastic two-body interactions occur in many systems, regimes where elastic 3-body interactions dominate are rare in nature. Here we discuss how the ubiquitous, though normally undesirable three-body losses of atomic physics experiments can induce effective three-body interactions. These are associated with interesting quantum phases, including Pfaffian states \cite{pfaffian}, and could be used to stabilise three-component Fermi mixtures \cite{threecomponent}, assisting in the production of a colour superfluid state \cite{coloursuperfluid}. We investigate Bosons in an optical lattice, where a three-body hard-core constraint stabilises the system with attractive two-body interactions, and a dimer superfluid phase emerges. We focus on the dynamics of this intrinsically time-dependent system, both testing the hard-core constraint for finite loss rates, and studying non-equilibrium properties including decay. In 1D, the exact evolution is computed by combining time-dependent density matrix renormalisation group methods (t-DMRG) \cite{vidal,tdmrg} with a quantum trajectories approach from quantum optics \cite{trajectories}. 

\begin{figure}[t]
\includegraphics[width=8.5cm]{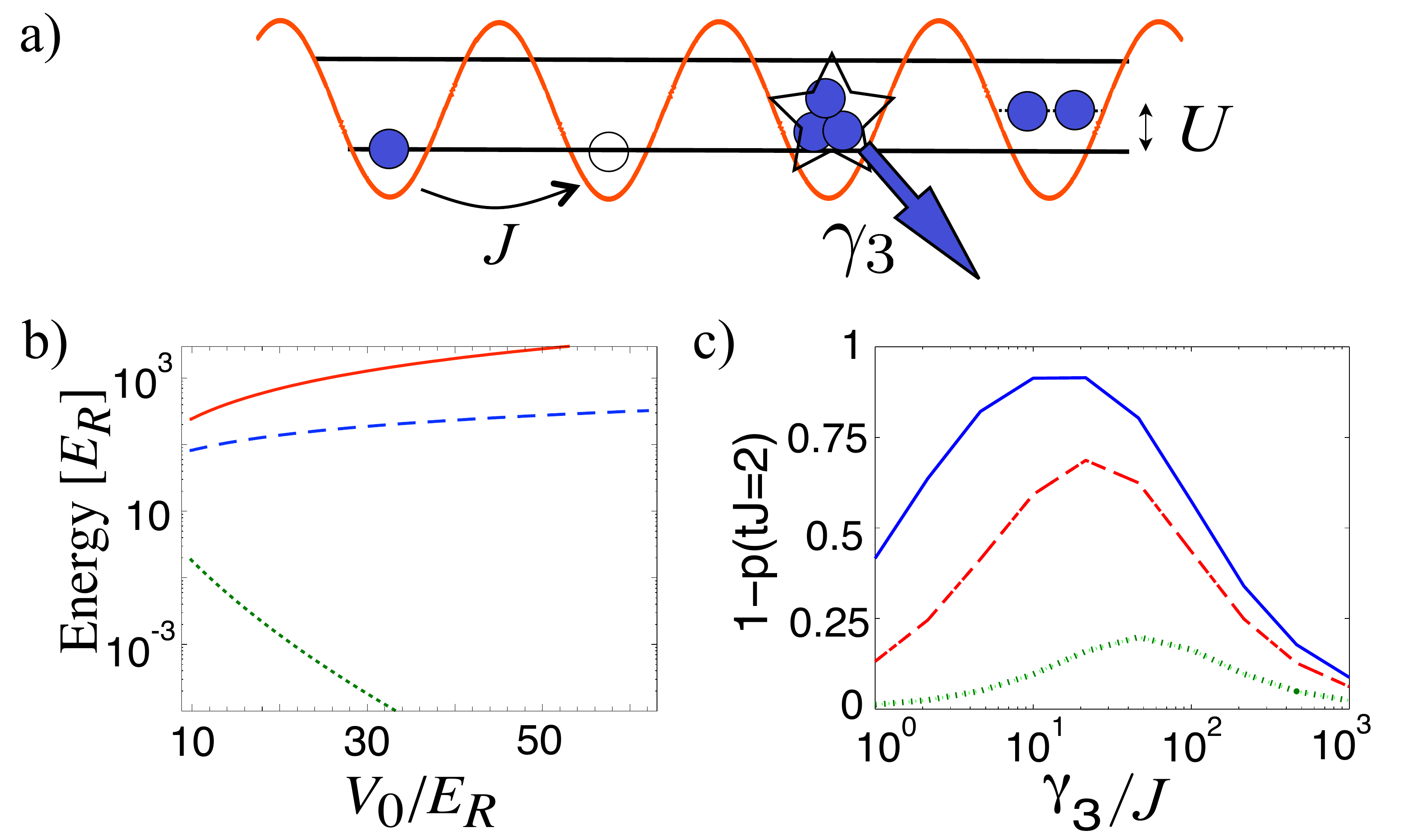}
\caption{(a) Bosons in an optical lattice in the presence of three-body loss at a rate $\gamma_3$. (b) Example model parameters estimated for Cs at a magnetic field of 15 Gauss (where the scattering length is $20 a_0$ and the recombination length $\sim 500 a_0$ \cite{csnumbers} where $a_0$ is the Bohr radius) as a function of lattice depth $V_0$, showing $\gamma_3$ (solid line), $U$ (dashed), and $J$ (dotted). Values of $\gamma_3$ are obtained by integrating the measured three-body recombination rates in free space over a state with three particles in a single Wannier function. (c) The probability that at least one loss event has occurred at time $tJ=2$, beginning with a single particle on each of 10 sites, and $U/J=3$ (solid line), 5 (dashed), and 10 (dotted), computed using t-DMRG methods (see text for details).
} \label{fig1}
\end{figure}

Three-body recombination \cite{csnumbers} in an optical lattice corresponds to decay into the continuum of unbound states, and thus loss from the lattice. This can be described by a master equation in the Markov approximation \cite{trajectories}, which for atoms in the lowest band of an optical lattice can be projected onto the corresponding basis of Wannier functions \cite{opticallattices, moleculeloss2}, associated with bosonic annihilation operators $b_i$ on site $i$. We can separate the master equation into terms which conserve particle number, corresponding to an effective Hamiltonian $H_{\rm eff}$, and terms which remove three particles on a site:
\begin{equation}
\dot \rho^{(n)} = - {\rm i}\left(H_{\rm eff} \rho ^{(n)}- \rho ^{(n)}H_{\rm eff}^\dag \right) +\frac{\gamma_3}{12} \sum_i 2  b_i^3 \rho^{(n+3)} ( b_i^\dag)^3, \nonumber
\end{equation}
where $\rho^{(n)}$ denotes the system density operator with $n$ atoms and $ \hat n_i= b_i^\dag  b_i$. The dominant loss term is on-site three-body decay \footnote{Off-site loss terms can arise from non-zero overlap of the corresponding Wannier functions, in analogy to off-site elastic interactions, which are discussed in L.-M. Duan, Europhys. Lett. \textbf{81}, 20001 (2008). These are small for $\gamma_3$ values used here.} and $\gamma_3$ is the corresponding rate. The effective Hamiltonian is
\begin{eqnarray}
H_{\rm eff}&=&H-{\rm i} \frac{\gamma_3}{12} \sum_i ( b_i^\dag)^3  b_i^3,\\
H &=& -J \sum_{\langle i,j\rangle } b_i^\dag  b_j + \frac{U}{2}\sum_i \hat{n}_i (\hat{n}_{i}-1) +\sum_i \varepsilon_i  \hat n_i, 
\label{eq:bhham}
\end{eqnarray}
with $J$ the nearest neighbor tunneling amplitude, $U$ the elastic two-body interaction, and $\varepsilon_i$ the local potential. The Hamiltonian is valid in the limit where $J,  \varepsilon_i, U n, \ll \omega$, with $\omega$ the band gap and $n$ the mean density. In an experiment these parameters, in particular $U$, may be tuned whilst $\gamma_3$ remains constant and large \footnote{This is because non-universal behavior (where $\gamma_3 \not\propto a^4$, with $a$ the scattering length) is observed away from a Feshbach resonance, where the rate of three-body recombination is large and weakly dependent on $a$ \cite{csnumbers}, while the two-body scattering length $a$ can be very small ($U\propto a$).}.  In Fig.~1c we show example values of $\gamma_3$, $U$, and $J$ using numbers for Caesium as a function of lattice depth.   

If we begin in a pure state with $N$ particles, then loss processes lead to heating, in that they produce a mixed state of different particle numbers. Within a fixed particle number sector, the dynamics are described by $H_{\rm eff}$. Three-body interactions emerge most clearly in the limit of rapid decay: $\gamma_3\gg J,U,\epsilon_i$. If we define the projector $P$ onto the subspace of states  with at most two atoms per site and $Q=1-P$, then in second order perturbation theory we obtain the effective model 
\begin{equation}
H_{\rm eff}^P \approx P H P - \frac{2{\rm i}}{\gamma_3} P H Q H P = P H P - {\rm i} \frac{6J^2}{\gamma_3} P\sum_j  c_j^{\dag} c_jP, \label{eq:model}
\end{equation}
where $ c_j = ( b_j^2/\sqrt{2}) \sum_{k\in N_j}  b_k$, and $N_j$ denotes the set of nearest neighbours of site $j$. The term $PHP$ describes the Hubbard dynamics, Eq.~(\ref{eq:bhham}), supplemented by the hard-core constraint $(b_i^\dag)^3=0$. Furthermore, the effective loss rates decrease as $J^2/\gamma_3$ \footnote{Note that off-site processes can place an upper bound on the useful value of $\gamma_3$, as increasing $\gamma_3$ suppresses on-site loss, but increases the rate of loss from processes involving neighbouring lattice sites.}.

Thus, we see the clear emergence of a three-body hard-core constraint in the limit $\gamma_3/J \gg 1$.  We can study the physics of the projected model $PHP$ to obtain a qualitative understanding of the quantum phases associated with the projection. However, the residual loss processes make this system intrinsically time-dependent, and can give rise to heating. We therefore study the full non-equilibrium dynamics, by combining t-DMRG methods \cite{vidal, tdmrg} with an expression of the master equation as an average over \textit{quantum trajectories} \cite{trajectories}. Each stochastic trajectory begins from an initial pure state (sampled from the initial density matrix), 
and can be interpreted as describing a single experimental run, in which losses occurred at particular times $t_n$ and on sites $i_n$. The evolution is described by the non-Hermitian $H_{\rm eff}$, except for times $t_n$, where losses (or \textit{quantum jumps}) occur, 
\begin{equation}
{\rm i}\hbar \frac{d}{dt}\ket{\psi(t)}=H_{\rm eff} \ket{\psi(t)}; \,\,\,\ket{\psi(t_n^+ )}=\frac{C_{i_n}\ket{\psi(t_n)}}{||C_{i_n}\ket{\psi(t_n)}||}, \label{trajevolution}
\end{equation}
where the jump operator $ C_i= b_i^3$ corresponds to three-body loss on site $i$. In stochastic simulation of the master equation, the times $t_n$ are points where the norm of the state falls below a randomly chosen threshold. At these times, a random jump operator is selected according to the probabilities $p_{i_n}\propto \bra{\psi(t_n)} C_i^\dag  C_i\ket{\psi(t_n)}$ and applied.  In this way we can both investigate individual trajectories and compute expectation values from the master equation. The latter is performed by stochastic average over both initial states and over jump events, which converges rapidly as the number of trajectories is increased.

The need to simulate many trajectories for convergence is offset by the efficiency of simulating states rather than density matrices. In contrast to directly applying t-DMRG to density matrices \cite{dissipativevidal}, we can simulate the master equation without squaring the local Hilbert space dimension, and by time-evolving states with fixed particle number, we also make use of existing optimisations for conserved quantities. Despite the application of local jump operators, we find the evolution quite efficient, especially for small numbers of jumps \footnote{We typically take the number of states retained in bipartite splittings \cite{vidal}, $\chi=200$ in the results presented here.}.  

As an example of the suppression of loss, we consider preparing a homogenous initial state at unit filling in a deep optical lattice where $U/J\rightarrow \infty$. At time $t=0$ we suddenly ramp the lattice to a finite depth, and observe the probability $p$ that a single three-body loss event has occurred as a function of time. In Fig.~1b we plot this probability for different $U/J$ as a function of $\gamma_3/J$. We see a clear suppression of loss rates for large $\gamma_3/J$, and also a substantial decrease for larger $U/J$, resulting from the decreased amplitude for doubly occupied sites. 

In the limit of large $\gamma_3$, it is instructive to study the equilibrium phase diagram of the projected Hamiltonian $PHP$. For $U/J>0$, we observe the well-known Mott Insulator (MI) and atomic superfluid phases of the Bose-Hubbard model. However, the three-body hard-core condition will also stabilise the system for $U/J<0$,  where we find a dimer superfluid phase (see Fig.~2a). This is characterised by the vanishing of the order parameter signalling superfluidity of single atoms (ASF) ($\langle  b_i \rangle=0$), while a dimer superfluidity (DSF) order parameter persists ($\langle  b_i^2 \rangle\neq 0$). The superfluid regimes are connected via a quantum phase transition associated with the spontaneous breaking of a discrete $Z_2$ symmetry, reminiscent of an Ising transition \cite{SR04}: the DSF order parameter transforms with the double phase $\sim \exp 2 \mathrm i \theta$ compared to the ASF order parameter $ \sim \exp \mathrm i \theta$. Consequently, the symmetry $\theta \to \theta + \pi$ exhibited by the DSF order parameter is broken when reaching the ASF phase.  

We can obtain a qualitative mean-field picture using a homogeneous Gutzwiller ansatz wavefunction, given for the projected Hilbert space by $| \Psi\rangle = \prod_i |\Psi\rangle_i$, where $|\Psi\rangle_i = r_0 {\rm e}^{{\rm i} \phi_0} |0\rangle_i + r_1 {\rm e}^{{\rm i} \phi_1} |1\rangle_i + r_2 {\rm e}^{{\rm i} \phi_2} |2\rangle_i$. Normalisation implies $\sum_\alpha r_\alpha^2 = 1$, while the filling is $n = r_1^2 + 2 r_2^2 \leq 2$.  
To examine the phases, we find the energy $E/M^d =  \langle \Psi | H |\Psi \rangle$, $E(r_\alpha, \phi_\alpha)=   U  r_{2}^2 - Jz r_1^2[  r_{0}^2 + 2 \sqrt{2}  r_{2} r_{0}   \cos\Phi + 2 r_{2}^2 ]$,
where $ \Phi = \phi_2+\phi_0 -2\phi_1$ and $M^d$ is the number of lattice sites. For any $r_\alpha$ the energy is minimized for $\Phi$ an integer multiple of $2\pi$. For $r_1, r_0\neq 0$ (placing us in the ASF phase with $|\langle  b_i \rangle|^2  = r_1^2 (r_{0} + \sqrt{2}r_{2})^2\neq 0$), the phase-locking expression contributes a ``source'' term linear in $r_2$ to the energy, and consequently the minimum of the energy cannot be located at $r_2=0$. Thus, a finite atomic condensate always implies a dimer component $|\langle  b_i^2 \rangle |^2 =  2 (r_{0}r_{2})^2$. However, the reverse is not true and one may have pure dimer superfluidity without an atomic condensate.   

At fixed $n$, the energy is a function, e.g., of $r_1$ alone. For the second order transition found within our mean field theory, an instability for atomic superfluidity is indicated by its mass term crossing zero, $\partial^2 E(r_1=0)/\partial r_1^2 \equiv 0$. This leads to a critical interaction strength for the ASF-DSF transition, $U_c/(Jz)  = - 2(1 + n/2 + 2\sqrt{n (1-n/2)})$.
Within the  DSF phase the order parameter obeys $|\langle b_i^2 \rangle|^2 = n(1-n/2)$ independent of the interaction strength. For $n\to 2$, we approach a MI state in a second order transition.  At $n=1$ we find that the ASF-DSF transition takes place at the same coupling strength as the ASF-MI transition, but with the opposite sign. The complete mean field phase diagram in the plane of density and interaction strength is plotted in Fig.~2a.

From the last term in Eq.~(\ref{eq:model}), we can estimate the initial loss rate from the ground state Gutzwiller wave function. We obtain the rate $\gamma_{\text{eff}}= 3 J^2 z/\gamma_3M^d(\langle \hat n_i^2\rangle - n)(n +| \langle \hat b_i\rangle|^2)$, which is zero in the MI, and $\propto n^2$ for the DSF, $\gamma_{\rm eff} = 3 J^2 zM^d n^2/\gamma_3$. In the DSF phase, the critical temperature $T_c \propto n^{2/3}$ at low densities, and the energy density deposited by a single loss, $\Delta E_{\text{loss}} =  (z+1)|U|n/(2M^d)$. The number of independent loss events needed to melt the DSF is then proportional to $T_c/\Delta E_{\text{loss}}$, and the melting time strongly decreases for increasing density, proportional to $\gamma_3/( |U|(z+1)J^2z n^{7/3})$.

These qualitative features are reproduced in 1D, as supported by numerical calculation of the ground state for $PHP$. In Figs.~2b,c we show the characterisation of the crossover between the ASF and DSF regimes in 1D via the off-diagonal elements of the single particle density matrix, $S(i,j)=\langle b_i^\dag b_j\rangle$, and the dimer density matrix $D(i,j)=\langle b_i^\dag b_i^\dag b_j b_j\rangle$. In the MI regime, the off-diagonal elements of $S(i,j)$ and $D(i,j)$ decay exponentially. As we enter the superfluid regime, quasi-long range order is visible in the polynomial decay (linear on the logarithmic scale). As $U/J$ is made more negative, we see a return to exponential decay for the off-diagonal elements of $S(i,j)$, but the off-diagonal elements of $D(i,j)$ still decay polynomially and, indeed, increase in magnitude. This characterises the DSF regime in 1D. Here, the transition to the DSF and MI regimes occurs at much smaller $|U/J|$ than in higher dimensions, but these two transitions again occur at similar $|U/J|$ for $n=1$.

\begin{figure}[tb]
\includegraphics[width=8.5cm]{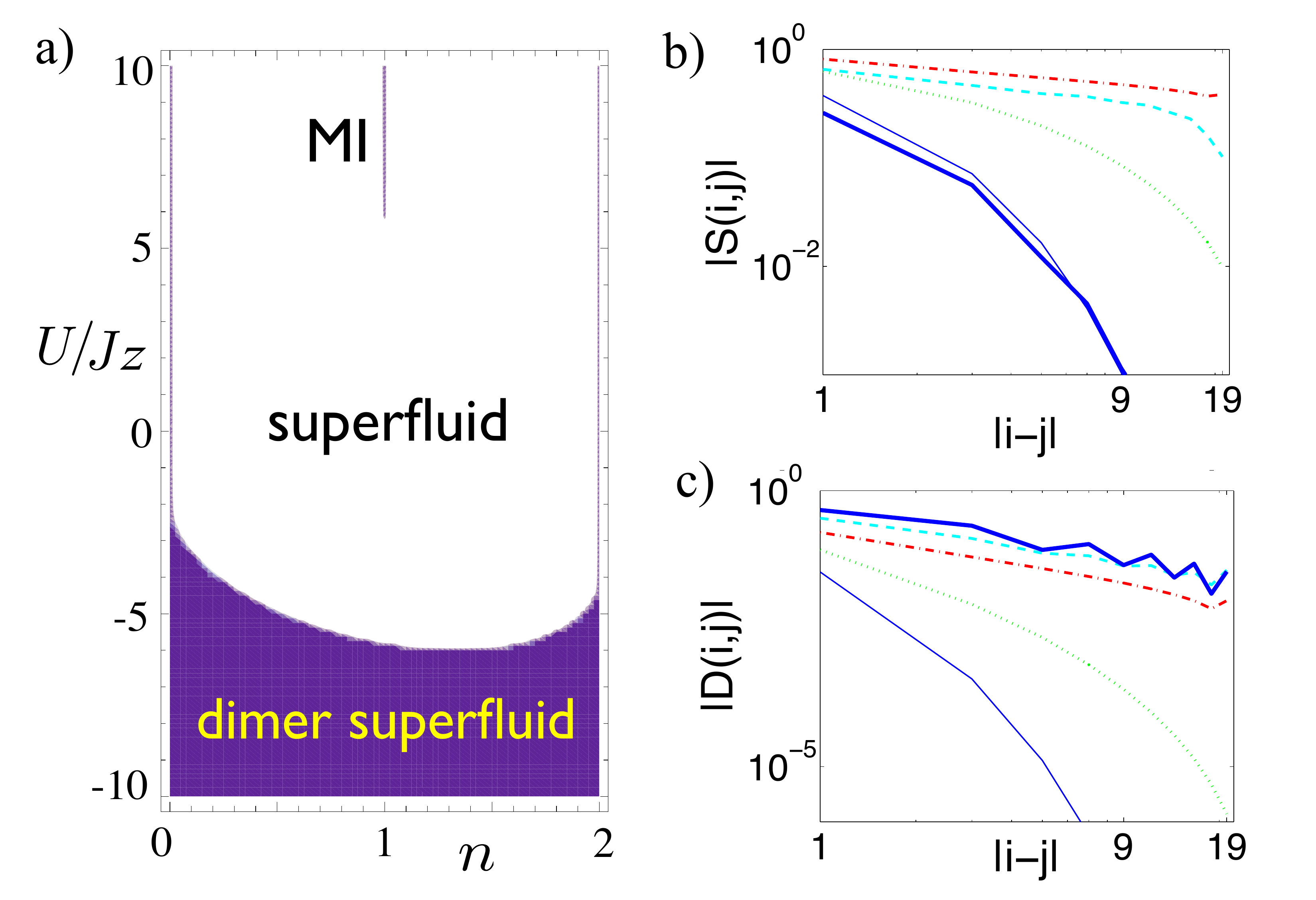}
\caption{Equilibrium analysis of the projected Bose-Hubbard model $PHP$. (a) Mean-field phase diagram as a function of $U/(Jz)$ and density, $n$. (b,c) Magnitude of off-diagonal elements of (b) the single particle density matrix  $|S(i,j)|=|\langle b_i^\dag b_j\rangle|$ and (c) the dimer density matrix, $|D(i,j)|=|\langle b_i^\dag b_i^\dag b_j b_j\rangle|$, as a function of $|i-j|$, for $U/J=$10 (thin solid line), 5 (dotted), 0 (dot-dash), -5 (dashed), and -10 (thick solid line). These results are computed for 20 particles on 20 lattice sites in 1D, with box boundary conditions and $i+j=21$, using imaginary time-evolution in t-DMRG, and plotted on a logarithmic scale.} \label{fig2}
\end{figure}

\begin{figure}[tb]
\includegraphics[width=8.5cm]{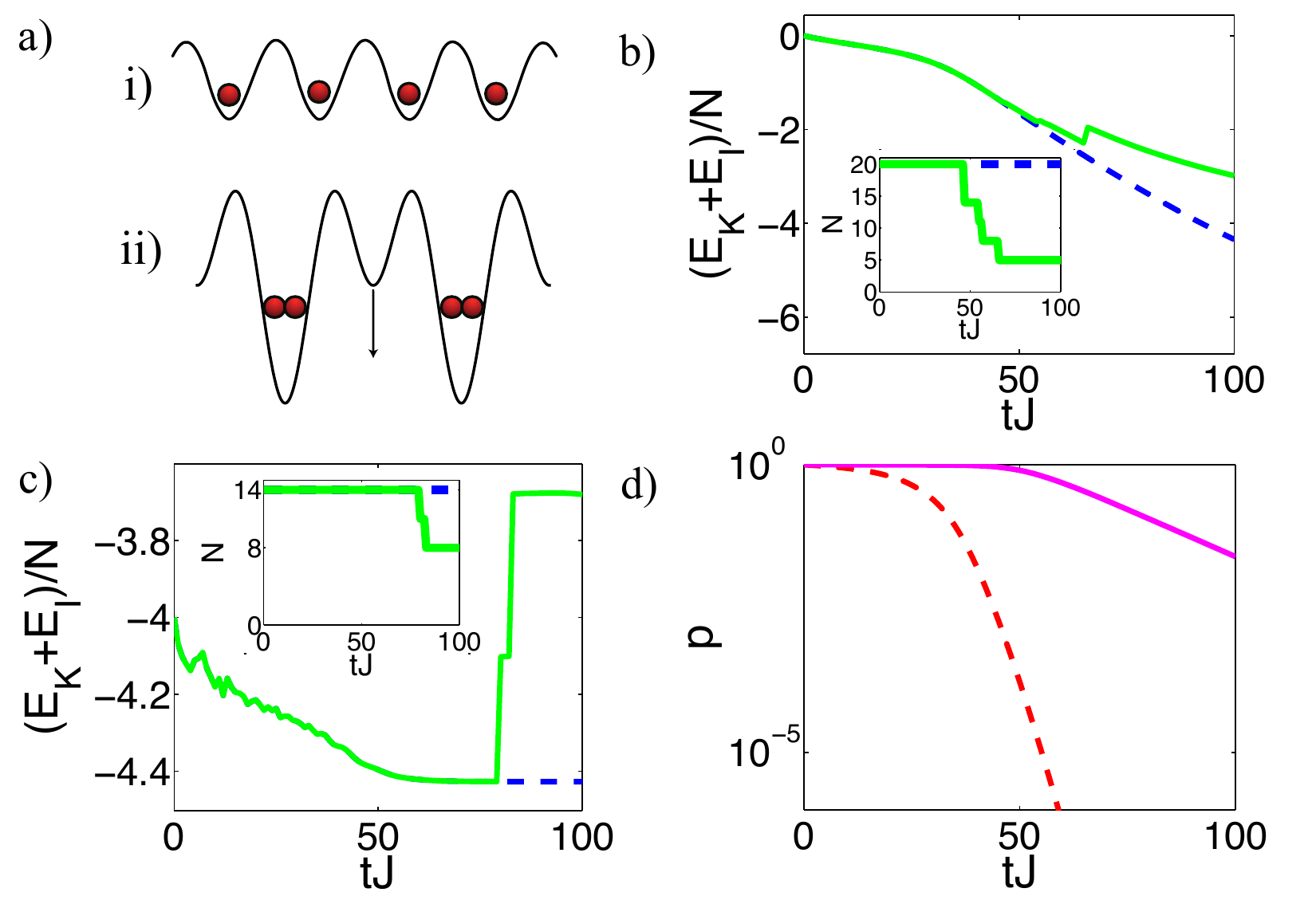}
\caption{Dynamics of adiabatic ramps into a dimer superfluid regime. (a) We begin with (i) a Mott-insulator state (ramping $U/J$), and (ii) a state with pre-prepared dimers in a superlattice (removing the superlattice). (b)-(c) The sum of kinetic ($E_K$) and interaction ($E_I$) energy and (inset) particle number as a function of time for two example trajectories, one with no loss events (dashed lines) and one with several loss events (solid lines). Here, (b) shows a ramp from $U/J=30$ to $U/J=-8$, with $U(t)=\alpha J/(100 + 3tJ)+\gamma$, with $\alpha$ and $\gamma$ ramp parameters, and (c) shows a ramp with a superlattice potential, $\varepsilon_l=V_0 \cos(2\pi l /3)$, where $V_0\approx30 J\exp(-0.1 tJ)$, adjusted so that $V_0(tJ=100)=0$, with fixed $U/J=-8$. In each case, $\gamma_3=250J$. For (b), we use 20 atoms on 20 lattice sites, for (c), 14 atoms on 23 lattice sites. (d) Plot showing the probability that no loss event has occurred after time $t$ for the ramps in (b) (dashed line) and (c) (solid).} \label{fig3}
\end{figure}
A dimer superfluid phase could be prepared in an experiment beginning in states that have very small amplitude of three-body occupation due to two-body interactions. Hamiltonian parameters could then be ramped adiabatically across the phase boundary, making use of dynamical suppression of three-body occupation. We study two such scenarios as illustrated in Fig.~3a: (i) We begin from a Mott Insulator state, and then ramp from $U/J=30$ to $U/J=-8$ in order to enter the dimer superfluid regime (which is intuitive, but associated with large probability of decay); or (ii) We apply a superlattice that raises the energy of certain sites and form a Mott Insulator with two particles per site in the lowest wells, then switch the interaction rapidly to $U/J=-8$ on a timescale much faster than tunnelling between the lowest wells and ramp down the superlattice. In each case, we compute dynamics using t-DMRG methods with quantum trajectories, and consider an on-site three-body loss with $\gamma_3/J=250$. We ramp parameters on a timescale where without loss the ground state will be reached with minimal presence of excited states.

In Figs.~3b,c we show the time dependence of the sum of kinetic and interaction energies, and of the total particle number for example trajectories. For each ramp type, we choose a ``lossless'' trajectory, where the ground state is reached adiabatically, and a ``lossy'' trajectory, where three-body loss events lead to heating of the system (as holes are produced that correspond to excited states).

In Fig.~3d, we compare the probability for each type of ramp of producing the ``lossless'' trajectory. 
We see that for the ramp from the MI state, where we pass through a region of small $U/J$, the probability of such a trajectory is essentially zero. For the superlattice ramp, on the other hand, it is much more likely that we will obtain the ground state from a randomly chosen trajectory. This is both because the superlattice allows us to use large $|U/J|$, so that the two-body interaction reduces the probability of triply occupied sites, and because the superlattice facilitates the choice of a lower density in the system, which significantly reduces the effective loss rate. 

In Fig.~4 we show the local density as a function of time and the final dimer density matrix $D(i,j)$ for (a) the ``lossless'' and (b) the ``lossy'' trajectories of Fig.~3c. When a loss occurs it affects not just the density for the site on which it occurs, but also on neighbouring sites due to the knowledge that we obtain of the position of the remaining particles. We also see clearly the destruction of correlations in the region of the system where the loss occurs. Note, however, that off-diagonal correlations are still visible in parts of the system, and that a single loss event does not always destroy the properties of the final state entirely. As discussed above, the probability of an individual loss event increases with system size, but a single loss event will change the character of the final state less.

\begin{figure}[tb]
\includegraphics[width=8.5cm]{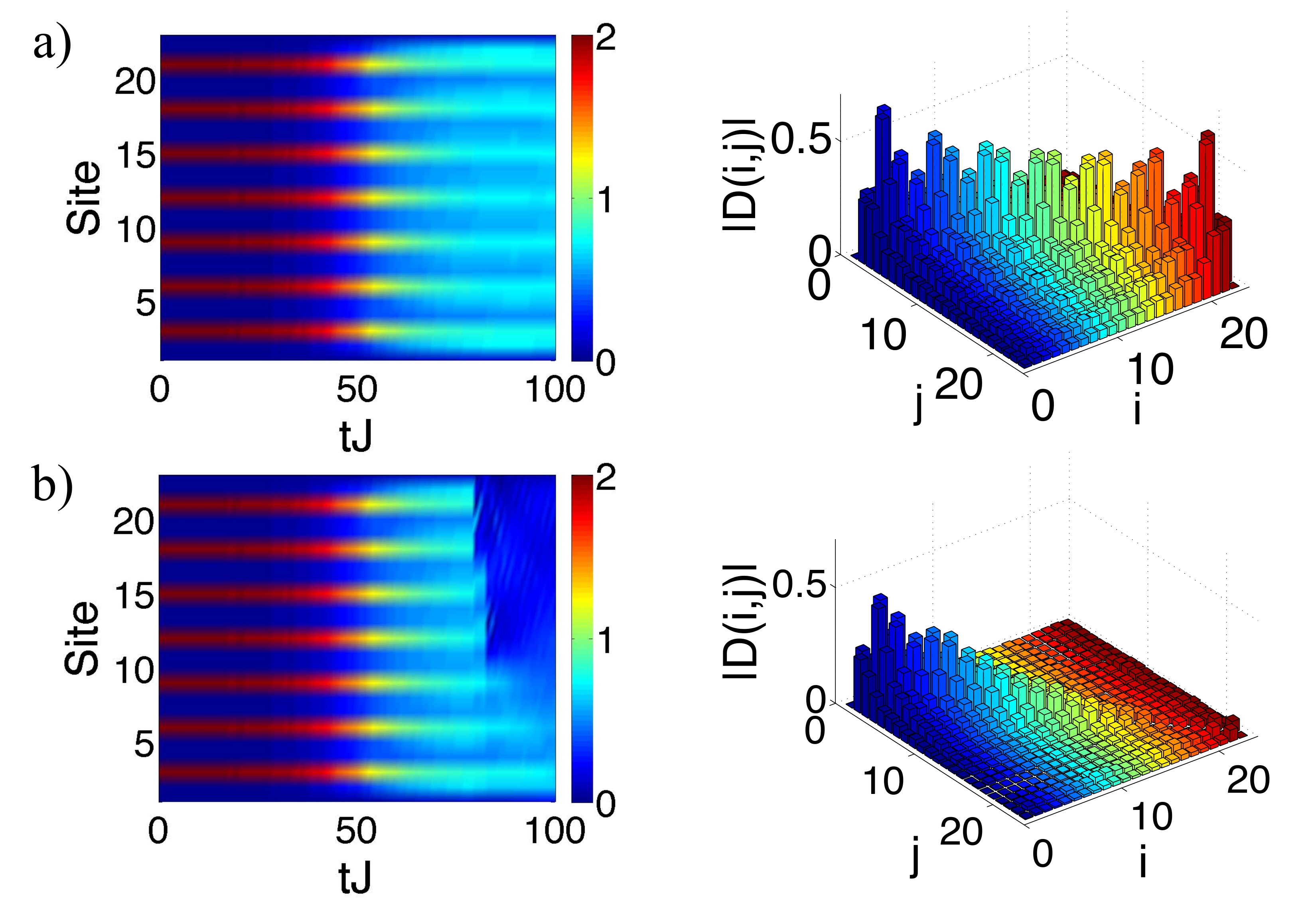}
\caption{Comparison of (a) lossless and (b) lossy trajectories from Fig.~3c. We show the mean density $\langle  n_i \rangle$ as a function of position and time, and magnitude of the dimer correlation function $|D(i,j)|$  ($i\neq j$) at the end of the ramp.} \label{fig4}
\end{figure}

The three-body interactions discussed here could have applications to producing Pfaffian-like states, and also for fermions, especially stabilising three-component mixtures. The combination of quantum trajectories methods with t-DMRG has potential applications in the simulation of other classes of master equations. In addition, there are open questions regarding the nature of the ASF-DSF phase transition, which we will address in more detail within a quantum field theoretical treatment of the attractive Bose-Hubbard model with a three-body hard-core constraint \cite{Diehl08}.

\begin{acknowledgements}
We thank L.-M.~Duan for discussions. Work in Innsbruck is supported by the Austrian Science Foundation (FWF) through SFB 15 and project I118\_N16 (EuroQUAM\_DQS), and by the DARPA OLE network. JMT is supported by a Pappalardo Fellowship at MIT.
\end{acknowledgements}

\end{document}